# Strain-Induced BCC-to-FCC Transformation in Tungsten Uncovered by Molecular Dynamics Simulations with Machine Learning Potentials

Qiu-Hong Hu,[1,2*] YueYue Wang,[2] Xinyi Leng,[2,3] Hezhu Shao,[4,†] Luru Dai,[2,3,‡].

[1]*The 2nd Affiliated Hospital and Yuying Children's Hospital of Wenzhou Medical University, Wenzhou, Zhejiang, 325027, China*

[2]*Wenzhou Institute, University of Chinese Academy of Sciences, Wenzhou, Zhejiang, 325000, China*

[3]*Postgraduate Training Base Alliance of Wenzhou Medical University, Wenzhou, Zhejiang, 325000, China*

[4]*Wenzhou Key Laboratory of Micro-nano Optoelectronic Devices, Zhejiang Key Laboratory of Smart Low-Voltage Apparatus and New Energy Application, College of Electrical and Electronic Engineering, Wenzhou University, Wenzhou, 325035, China*

Polymorphic transformations occur in a minority of metals. Tungsten is not one of them. This observation does not rule out its intrinsic phase-transition possibility, but may be due to inaccessible thermodynamic and mechanical conditions in conventional experiments. Herein, we investigate the body-centered cubic (BCC) to face-centered cubic (FCC) transformation in tungsten via a rigorously validated machine learning potential (MLP) molecular dynamics (MD) approach. Our simulations demonstrate that strain-induced BCC to FCC transformation indeed occurs in tungsten. The FCC phase exceeds 20 at.% at low temperature and in the presence of dislocations. In the perfect crystal moderate temperatures favor phase transformation kinetics. Structural analysis confirms the transformation following the Bain and Nishiyama-Wassermann orientation relationship. These results establish the viability of strain-induced polymorphic transformation in tungsten and provide a critical guideline for the experimental production of duplex tungsten with enhanced ductility—addressing a longstanding challenge in tungsten material engineering for high-temperature structural applications.

In the close-packed structure of crystalline metals, FCC and hexagonal close-packed (HCP) structures achieve a space filling of 74% through ABCABC or ABAB stacking sequences, respectively [1, 2]. These structures feature 12 nearest neighbors and 12 independent slip systems, which correspond to the geometric limit of maximum atomic close-packing in three-dimensional lattices. In contrast, the BCC structure features a lower space filling of 68% with eight nearest neighbors and only 8 independent slip systems, yet maintains thermodynamic stability at room temperature [3]. The more the slip systems, the higher dislocation mobility, leading to significantly better ductility and toughness. Metals with a BCC structure, such as tungsten, molybdenum, tantalum, and niobium are known for their exceptional properties including high melting point, low sputtering yield, and excellent high-temperature strength required for applications in extreme environments [4-11]. However, their inherent brittleness significantly limits their broader usage as structural materials [12]. Recent experiments showed that tungsten can form a transient FCC-like atomic packing at the tip of cracks, with the spatial extension slightly above the atomic scale during plastic deformation at high temperatures [13], indicating a slim possibility for obtaining close-packed structures from metals with a BCC structure [14-17].

A conventional mechanical test as such is unable to capture atomic-scale dynamic processes, like dislocation nucleation and slip band interactions [18, 19]. Classical MD simulations can assess dynamic processes on microsecond timescales, but their applications are limited by the inaccuracy of empirical interatomic potentials in describing complex defect structures [20]. Density Functional Theory (DFT) offers precise descriptions of electronic structures, its high computational cost for large systems limits applications in modeling large-volume lattice distortions during plastic deformation [21, 22]. MLPs through local atomic environment descriptors combine the accuracy of DFT with the computational efficiency of classical MD simulations [23-26]. These potentials have demonstrated notable advantages in the studies of phase transformation in titanium alloys [36] and the strengthening-toughening mechanisms of high-entropy alloys [37]. Recent efforts in the development of MLPs for tungsten incorporated data on radiation damage and lattice dynamics [38-40]. However, their predictive power for non-equilibrium processes such as multi-slip system competition and dislocation motion has not yet been validated.

Herein, we combine DFT with MLPs to validate MLP-based GPUMD simulations of tensile loading in single-crystalline BCC tungsten, with temperature and pre-introduced dislocations as key variables, and characterize the deformation behavior via stress-strain curves. We found that pronounced ductility (15.7%) is achieved at 1 K with deformation dominated solely by the {110} slip system. {110} remains the dominant slip plane at 300 K and 2000 K, accompanied by activated ( $33\bar{1}$ ) and ( $55\bar{3}$ ) planes for plastic deformation. Such plastic deformation through multi-slip systems was experimentally reported by Cecile Marichal and C.R. Weinberger [41, 42]. We observed strain induced BCC-FCC phase transformation. The amount of FCC phase was determined by Common Neighbor Analysis (CNA) and Polyhedral Template Matching (PTM) in OVITO [43-46]. The maximum content of FCC phase reached above 21 at.%. The transformation proceeds along the Bain and Nishiyama-Wasserman path [47-50] via a three-step structural evolution: BCC → BCT→ monoclinic → FCC. The presence of edge ($\bar{b} \parallel z$, $\bar{b} \parallel$ x) and screw ($\bar{b} \parallel x$, $\bar{b} \parallel z$) dislocations reduces the threshold of strain for this BCC-FCC transformation, with edge-$\bar{b} \parallel x$ showing the most pronounced effect [51].

The knowledge gained from these findings can be applied to other refractory metals and their alloys. The realization of FCC phase via strain engineering represents a viable strategy to overcome intrinsic brittleness, enhance deformability, and achieve balanced strength-toughness performance for extreme-environment applications.

**Adaptation and Validation of the NEP Model**

Extended efforts have been made to apply the NEP model in the study of metals and alloys. Recently, Keke Song et al [52] have developed the Unified Neuro Evolution Potential (UNEP-v1) for general-purpose MLPs in materials science. And 16 metals and their alloys are tested as model systems. For each unary and binary system, an initial training set is constructed comprising hundreds of structures. This dataset contains a broad range of atomic configurations, including defect structures such as vacancies, dislocations and stacking faults for single-crystals; and other defects like grain boundaries for polycrystal materials. Other conditions like temperatures and stresses are also included.

We extracted the elemental tungsten structures from the UNEP-v1 potential and designated this dataset UNEP-v1_w. In addition, to evaluate the generalization capability of the model, we employed the UNEP-v1 potential to generate 90 new tungsten configurations, denoted as NEP_w. For the uniaxial tension simulations based on DFT using the Perdew–Burke–Ernzerhof (PBE) functional, we choose the BCC-tungsten supercell as the test sample and apply uniaxial tension along the [100] direction, with each deformation increasing by a strain of 1%. For the initial state, i.e., when the strain is 0, we release the constraints in all three principal axis directions ([100], [010], and [001]) and conduct structural optimization. For the structures with strain, we only fix the degrees of freedom in the tensile direction while optimizing the remaining directions. Finally, 30 structures are obtained. For the uniaxial tension based on MD simulations, four target temperatures are considered: 1K (This temperature only serves to suppress thermal fluctuation, which is approximately equivalent to an effective temperature of 0 K within the framework of DFT), 300 K, 1000 K, and 2000K (Tungsten has a melting point of 3695 K) For each temperature, the BCC- tungsten supercell is subjected to uniaxial tension along the [100] direction at a strain rate of $6.286\times10^{9}$/s until fracture. By extracting 15 structures from each uniaxial tension simulation, a total of 60 structures is obtained.

Principal component analysis (PCA) using the pynep package [53] shows a considerable overlap in the feature space between the UNEP-v1_w and the

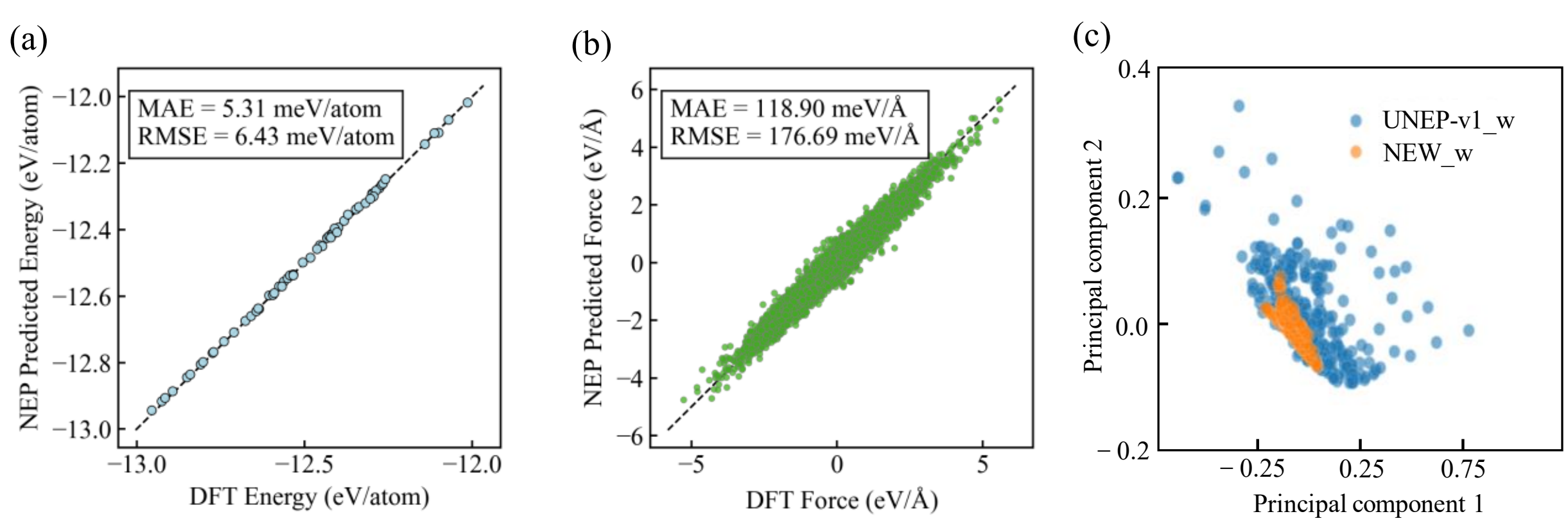

*qiuhong.hu@ucas.ac.cn
†hzshao@wzu.edu.cn
‡dai@ucas.ac.cn

FIG1. (a) The overall energy Mean Absolute Error (MAE) of the model predicted using UNEP-v1. (b) The overall force MAE of the model predicted using UNEP-v1. (c) PCA Analysis in Feature Space for UNEP-v1_w and NEP_w

NEP_w (Fig. 1). This indicates that the initial training set is sufficient for a comprehensive representation of the configurational space of these materials in our applications. The baseline NEP model exhibits outstanding accuracy on these out-of-sample multi-temperature tensile configurations without any need for further retraining. Specifically, the overall Energy Mean Absolute Error (MAE) is only 5.31 meV/atom, and the Force MAE is 119 meV/Å. These results firmly validate that UNEP-v1 model possesses excellent generalization capability to capture the complex thermal fluctuations and lattice deformations up to 2000 K. Therefore, molecular dynamics simulations in this study are performed using the UNEP-v1 potential.

To further evaluate the performance of the UNEP-v1 model under various conditions, we also examine its ability to predict phonon dispersion relations at different temperatures. The phonon dispersion relations of BCC-W calculated from the UNEP-v1 model are in good agreement with experimental data at 22℃ [54] and previous DFT calculations [55](Fig. 2(a)). Noticeably, as the temperature rises, manifesting as a monotonic shift of phonon frequencies toward lower values (Fig. 2(a)). This phenomenon plays a key role in understanding the temperature-dependent mechanical properties of the materials.

Given that elastic constants can be analytically derived from the second-order interatomic force constants (IFCs) using Born's long wave method [56], the MD+DynaPhoPy approach is an effective framework for evaluating the elastic properties of solids at finite temperatures. Fig. 2(b) illustrates the temperature dependence of the calculated elastic constants, which shows an overall good agreement with existing experimental data [57, 58]. The results reveal a gradual degradation of mechanical stiffness with increasing temperature. According to the Born criteria [17], the dynamic stability of a cubic crystal is ensured, if the following conditions are satisfied: $c_{11} + 2c_{12} > 0, c_{11} - |c_{12}| > 0$ , and $c_{44} > 0$ . Within the temperature range, the crystal remains stable. However, the increase in temperature causes instability accompanied by a decrease in elastic constants. Near the melting point, the stability of the crystal is nearly destroyed. For BCC-W, $c_{11}$ is particularly temperature-sensitive, decreasing by approximately 17% from 300 K to 1800 K in the experimental study of elastic properties of tungsten [57], and our simulation gives a reduction of 16.4%. In contrast, $c_{12}$ and $c_{44}$ exhibit minimal temperature dependence.

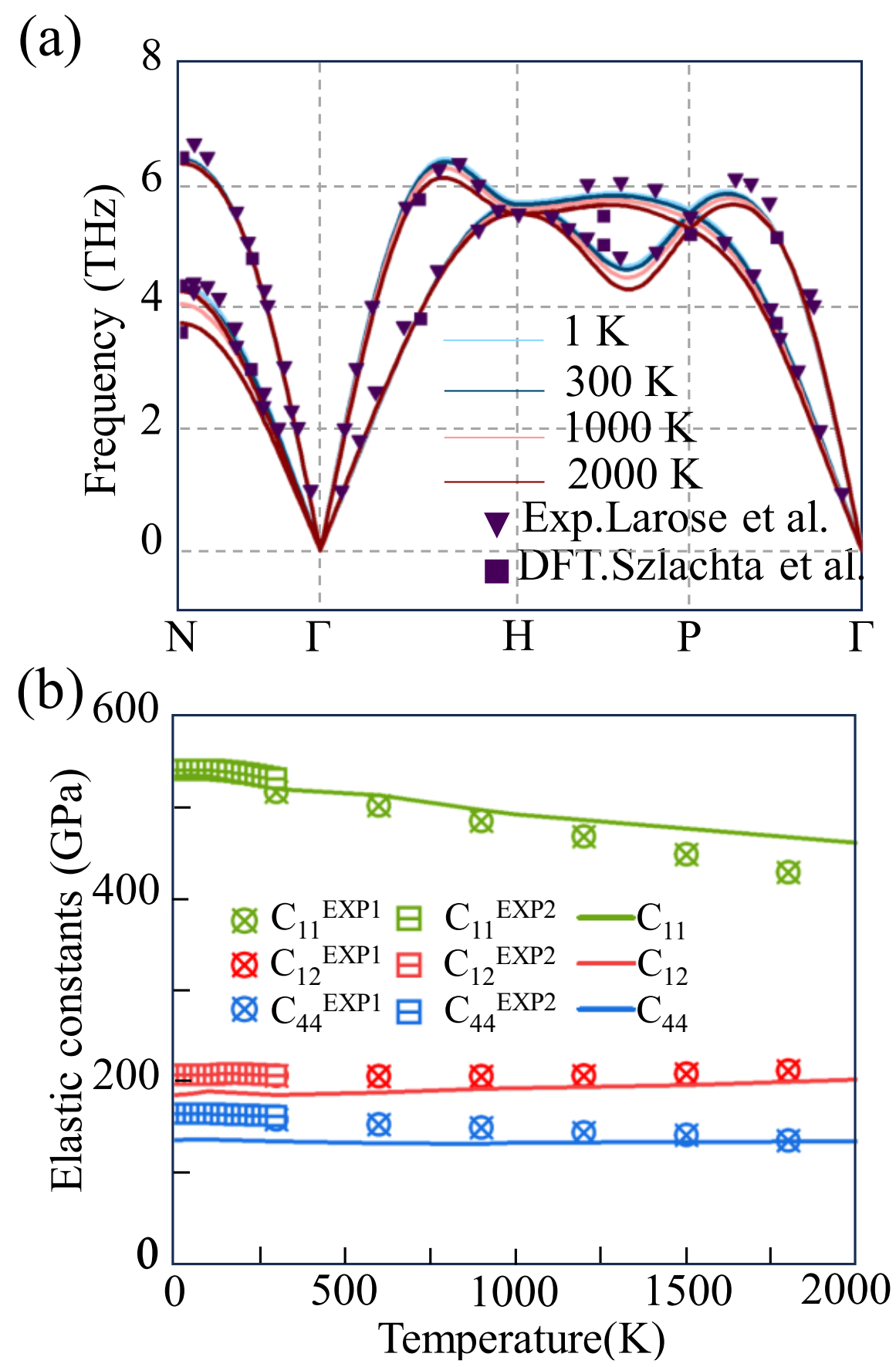

FIG 2. (a) Phonon dispersion relations of tungsten (1-atom primitive cell) from UNEP-v1 at various temperatures, compared with DFT calculations [55] and experimental data at 22 ℃ [54], along high-symmetry directions in the Brillouin zone. (b) Elastic constants $C_{ij}$ (GPa) of BCC tungsten, calculated via MD+DynaPhoPy from finite-temperature second-order IFCs, versus experimental values [57, 58]

**MD simulations on the mechanical behavior of tungsten**

Systematic MD simulations are conducted using the UNEP-v1 model with GPUMD software (version 3.9.5) [53]. Five different single-crystal tungsten models, including one perfect crystal and four crystals with dislocations, are generated using the ATOMSK package [51], we introduce edge dislocations with Burgers vector $\bar{b} \parallel z$ (edge-$\bar{b} \parallel z$) and $\bar{b} \parallel x$ (edge-

*qiuhong.hu@ucas.ac.cn
†hzshao@wzu.edu.cn
‡dai@ucas.ac.cn

$\bar{b} \parallel x$ ), respectively; and screw dislocations with Burgers vector $\bar{b} \parallel x$ (screw-$\bar{b} \parallel x$) and $\bar{b} \parallel z$ (screw-$\bar{b} \parallel z$). The corresponding dislocation densities are 0.049% and 0.098%, respectively. All models are approximately 31-32 nm in length, with widths and heights of 6-7 nm. Uniaxial tensile simulations are performed on the constructed models at four different temperatures: 1 K, 300 K, 1000 K, and 2000 K. The system is first relaxed in the isothermal-isobaric ($NpT$) ensemble at the target temperature using the Berendsen thermostat [59] for 20 ps. Subsequently, the model is stretched in the $x$ direction at a strain rate of $6.286\times10^{9}$/s. We used the Dislocation analysis (DXA) function and the CNA tool in the OVITO software to track the dislocation structure and local crystal configuration, respectively [60]

**Tensile process from elastic deformation through phase transformation to fracture**

Fig. 3 illustrates the stress-strain relations of single-crystal tungsten at different temperatures. The curves can be divided into four intervals: (1) quasi-elastic deformation ($O_1 \rightarrow A_1, O_2 \rightarrow A_2, O_3 \rightarrow A_3, O_4 \rightarrow A_4$), (2) yielding ( $A_1 \rightarrow B_1$ ), (3) BCC → FCC phase transformation ( $B_1 \rightarrow C_1, A_2 \rightarrow B_2, A_3 \rightarrow B_3, A_4 \rightarrow B_4$ ), and (4) fracture ( $B_1 \rightarrow C_1, B_2 \rightarrow C_2, B_3 \rightarrow C_3, B_4 \rightarrow C_4$). The subscript "1" denotes 1 K, while "2", "3", and "4" denote 300 K, 1000 K, and 2000 K, respectively.

*Quasi-elastic deformation:* During the initial stage of tensile straining, the atomic spacing along the loading direction gradually increases with applied strain in a manner that can be approximated by a quadratic function as shown in Fig. 3(b)-(f). When the stress is unloaded under 1 K, the system reverts to its initial state (Fig. 4(a)). This behavior is observed under both low, moderate and high temperatures, which suggests that the deformation mechanism in this stage is quasi-elastic.

*Yielding:* At the temperature of 1 K, the material yields when the stress reaches the peak at point $A_1$. No phase transformation or dislocations are generated at this stage, and yielding merely originates from the variation of atomic spacing. As stress further increases, the system enters plastic instability. If the load is removed within this stage, the material cannot revert to its initial configuration (Fig. 3(b)). At the temperatures of 300 K, 1000 K and 2000 K (Fig. 3(d)-(f)), no yield is present. This can be attributed to enhanced thermal activation effects, which increase atomic mobility, thereby confirming the dominant role of the thermal activation mechanism in plastic deformation [62].

*Phase transformation:* From the analysis of stress-strain curves and CNA, the phase transformation takes place at the strain interval of $B_1 \rightarrow C_1$. The phase transformation accompanies a sharp drop in stress and crack generation at 1 K (Fig. 3(c)). At the temperatures of 300 K, 1000 K and 2000 K (Fig. 3(d)-(f)), in contrast, the phase transformation corresponds only a slight decrease in stress. No cracks observed. The difference in the mechanical response concurrent

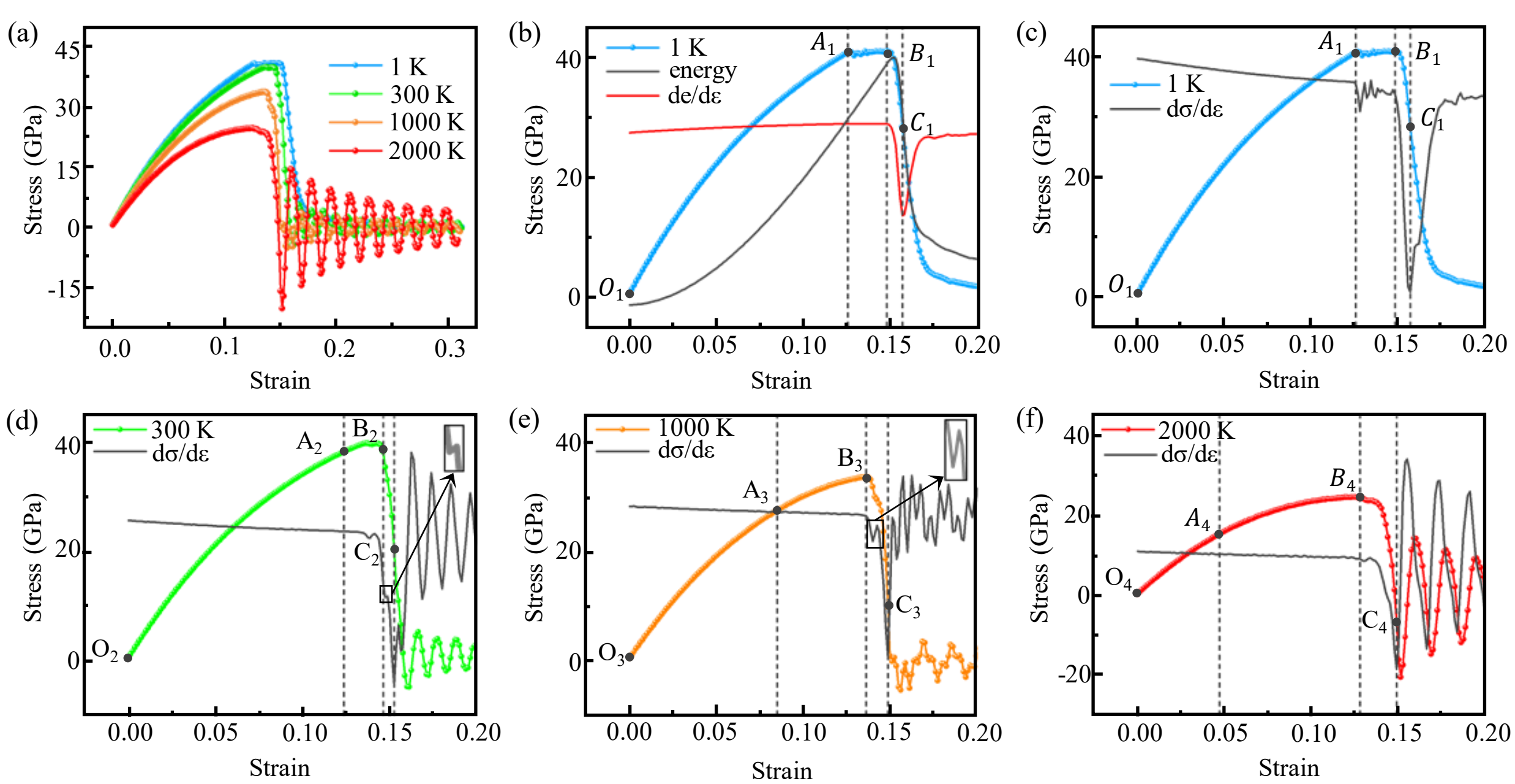

*qiuhong.hu@ucas.ac.cn
†hzshao@wzu.edu.cn
‡dai@ucas.ac.cn

FIG 3. (a) Stress-strain curves of perfect single-crystals at various temperatures (b), Stress-strain curve, total energy, and de/dε curve at 1 K. (c), (d), (e), (f), Stress-strain curve and dσ/dε curve at 1 K (c), 300 K (d), 1000 K (e), and 2000 K

fracture damage occurring in the $B_1 \rightarrow C_1$ interval at 1 K, and absence of cracks at other temperatures. At 1 K, the BCC→FCC phase transformation is primarily initiated through the {110} slip planes (FigS. 1), less on higher-index slip planes such as $(11\bar{3})$ and $(55\bar{3})$ (FigS. 1(g)). Slip planes acting as stress concentration regions allow stress to induce lattice distortion in their vicinity and create conditions for phase transformation. In addition, a large number of $\frac{a}{2}\langle 111\rangle$ dislocations are generated during this stage at 300 K (FigS. 2(d)), resulting in a temporary increase in the slope of the stress-strain curve $(d\sigma/d\varepsilon)$, as indicated by an enlarged rectangle at the head of the arrow in Fig. 3(d). The kink in the curve can be seen as a reflection of the localized hardening by dislocation multiplication [63]. The coincidence sites lattice (CSL) [64] analysis of the sample volume suggests the appearance of Σ3<111> and Σ29 boundaries on the slip planes. At low temperatures, only very few Σ29 and Σ3 boundaries are observed. They occurred predominantly on the {110} slip planes (FigS. 1). At 300 K, more Σ3 and Σ29 boundaries are observed, primarily on the {110} and higher-index slip planes (FigS. 2). Similar tensile behaviors during this stage are observed at 1000 K and 2000 K. According to the energy-strain curve and its first derivative during the tensile process at 1 K (Fig. 3(b)), the energy initially increases gradually until reaching a maximum value, while the rate of energy increase diminishes to zero. Further stress application leads to a negative energy growth rate, followed by a sharp decline in energy. Throughout this process, the extent of the BCC→FCC transformation continuously increases, indicating that the phase transformation is accompanied by energy dissipation.

*Fracture:* When the strain reaches point $B$, further loading leads to a rapid decrease in stress, and finally to the fracture of the material. The highest peak point of the stress curve is defined as the Ultimate Tensile Strength (UTS). At 1 K, cracks initiate and propagate along the BCC→FCC phase boundaries. At the temperatures of 300 K, 1000 K and 2000 K, the convergence of multiple slip planes induces localized stress concentration, ultimately leading to material failure.

In addition, to verify the reversibility of deformation within the elastic range, we removed the applied stress at different deformation stages. The results indicate that after unloading at various stress values, the system can fully relax and recover to its initial state in the quasi-elastic stage, whereas it cannot return to the initial state during the phase transformation and fracture stages (Fig. 4(a)(b)(c)). We also perform simulations on the effect of different tensile rates on the stress–strain curve at 1 K (Fig. 4(d)). The results indicate that the peak tensile strain and stress gradually increase with the rise of tensile rate, which is mainly attributed to the massive accumulation of lattice distortion and the slow response of the materials.

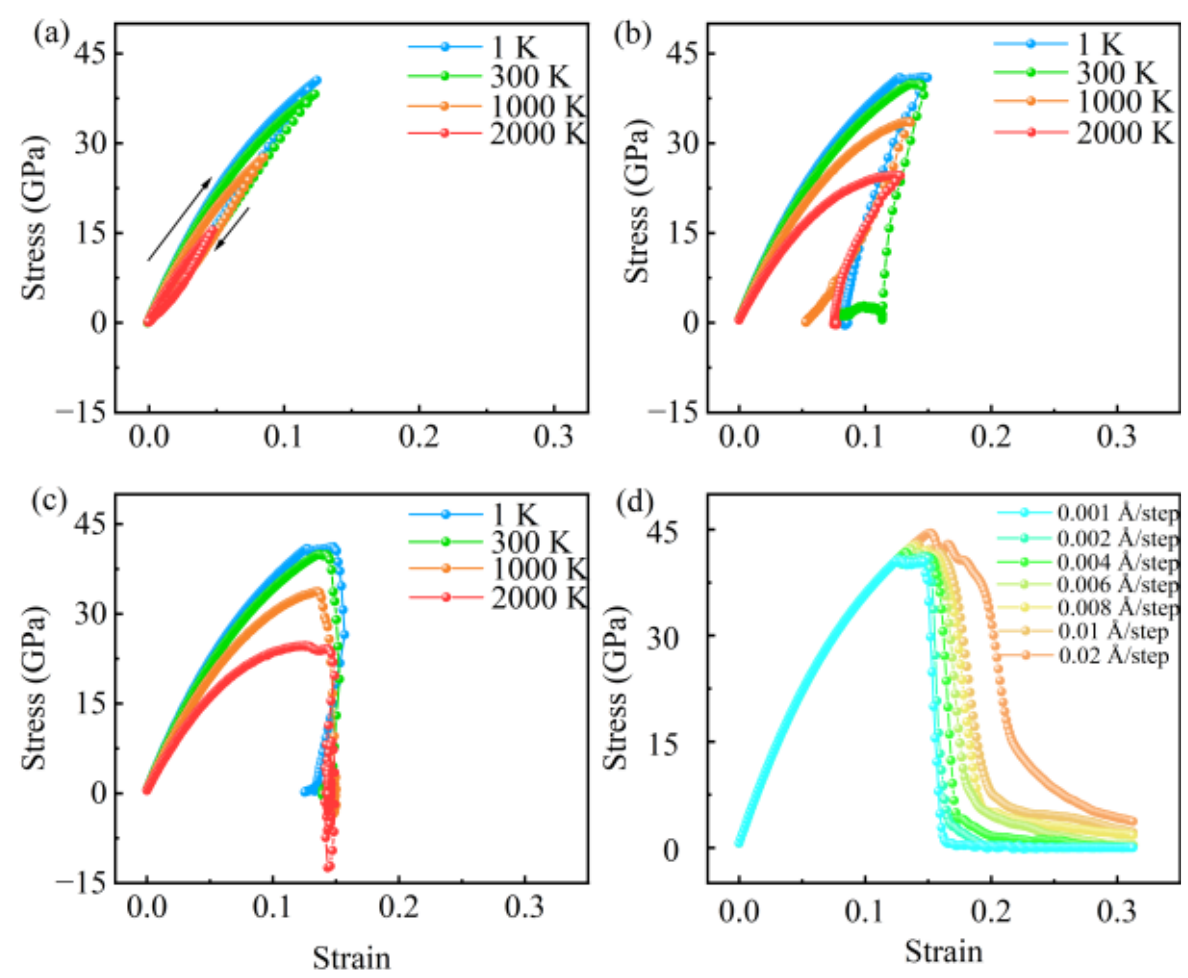

FIG 4. (a) Structural relaxation after stress unloading in the quasi-elastic deformation stage; (b) Structural relaxation after stress unloading in the phase transformation stage; (c) Structural relaxation after stress unloading in the fracture stage; (d) Uniaxial tensile stress–strain curves of single-crystal tungsten at different strain rates under 1 K.

Furthermore, in order to test the theoretical robustness and the consistency with the experimental results, we analyzed tensile stress data found in the literature [8, 64, 66]. In spite of the differences in testing conditions and sample sizes in these studies, we extracted the critical mechanical parameters and conditions, including UTS, temperatures and tensile orientation. The UTS calculated by our GPUMD simulations use the UNEP-v1 potential. The calculated UTS in this work agrees well with the results of molecular dynamics (MD, 15–47 GPa) [55] and density functional theory (DFT, 29–54 GPa) [56] simulations. On the other hand, the bulk tensile experimental data show one order of magnitude lower in UTS (< 1.2 GPa) [8]. Such a low experimental value is generally considered to stem from the high density of crystal

* qiuhong.hu@ucas.ac.cn
† hzshao@wzu.edu.cn
‡ dai@ucas.ac.cn

defects inside the as-fabricated single-crystal tungsten specimens for testing.

To summarize the results of the tensile process: (1) The mechanical behavior of perfect crystal under uniaxial tension is strongly dependent on temperature. As the temperature increases, the yield phenomenon disappears, and both the stress and strain corresponding to the phase transformation point progressively decrease. (2) BCC→FCC transformation takes place during the tensile process, and the amount of phase transformation is greatest at 1000 K (10.35%) (FigS. 3(c)). However, FCC phase transforms back into the BCC phase as cracks initiate. (3) the main slip plane remains {110} in a wide temperature range, and (4) the structure of the fracture region resembles that of Σ3 and Σ29 boundaries.

**The path of BCC→FCC transformation**

Fig. 5(a)(b)(c) display the local atomic configuration maps of the perfect crystal under uniaxial tension at 1 K with a strain of $\varepsilon$ = 14.9%, where it can be seen that some atoms of the $(0\bar{1}1)_{BCC}$ plane transform into an BCT structure. Aoltoms on the BCC (100) plane contract along $[011]_{BCC}$ and $[01\bar{1}]_{BCC}$ (Fig. 5(a)), while expanding along $[100]_{BCC}$ and $[00\bar{1}]_{BCC}$ (Fig. 5(b)). Fig. 5(d) is a diagram of the BCC→BCT transformation. The black atoms are those in a BCC unit cell that are about to deform, and the grey atoms are the atoms after the black atoms expand or contract along the arrow directions. The initial lattice distortion from BCC to BCT resembles the Bain path but differs in the ratio of contraction/expansion compared to the original Bain model. At $\varepsilon$ = 15.1%, the atoms on the $(110)_{BCC}$ plane shear along the $[\bar{1}01]_{BCC}$ and $[10\bar{1}]_{BCC}$ direction (Fig. 5(f)). This process deforms the BCT structure into a monoclinic structure (Fig. 5(e)). Fig. 5(h) is a diagram of the BCT → monoclinic transformation. The grey atoms shear along the grey arrows to the position of the red atoms, forming a monoclinic structure with the four unsheared grey atoms. At a strain of ε = 15.2%, the monoclinic structure undergoes continuous shear deformation and orthorhombic lattice distortion along the Nishiyama-Wassermann path, eventually completing the transformation into the FCC phase. This represents a typical martensitic transformation process [38, 57].

The bond-angle relationships can be seen of the BCC, BCT, monoclinic, and atoms on $(101)_{BCC}$ plane (Fig. 5(c)(g)(k)). Before $\varepsilon$ = 14.9%-15.1%, no cracks are generated, and the bond angles of BCC → BCT → monoclinic gradually increase with the continuous increase of strain. When $\varepsilon$ = 15.2%, cracks begin to appear, and the corresponding stresses of BCC and FCC near the cracks are released. At this strain value, the two phases are both stable as clearly reflected by relationships between monoclinic → FCC phases on (100), $(0\bar{1}1)$, and (101) planes, respectively. (l) Monoclinic→ FCC transformation via an orthorhombic intermediate at ε = 15.2%.

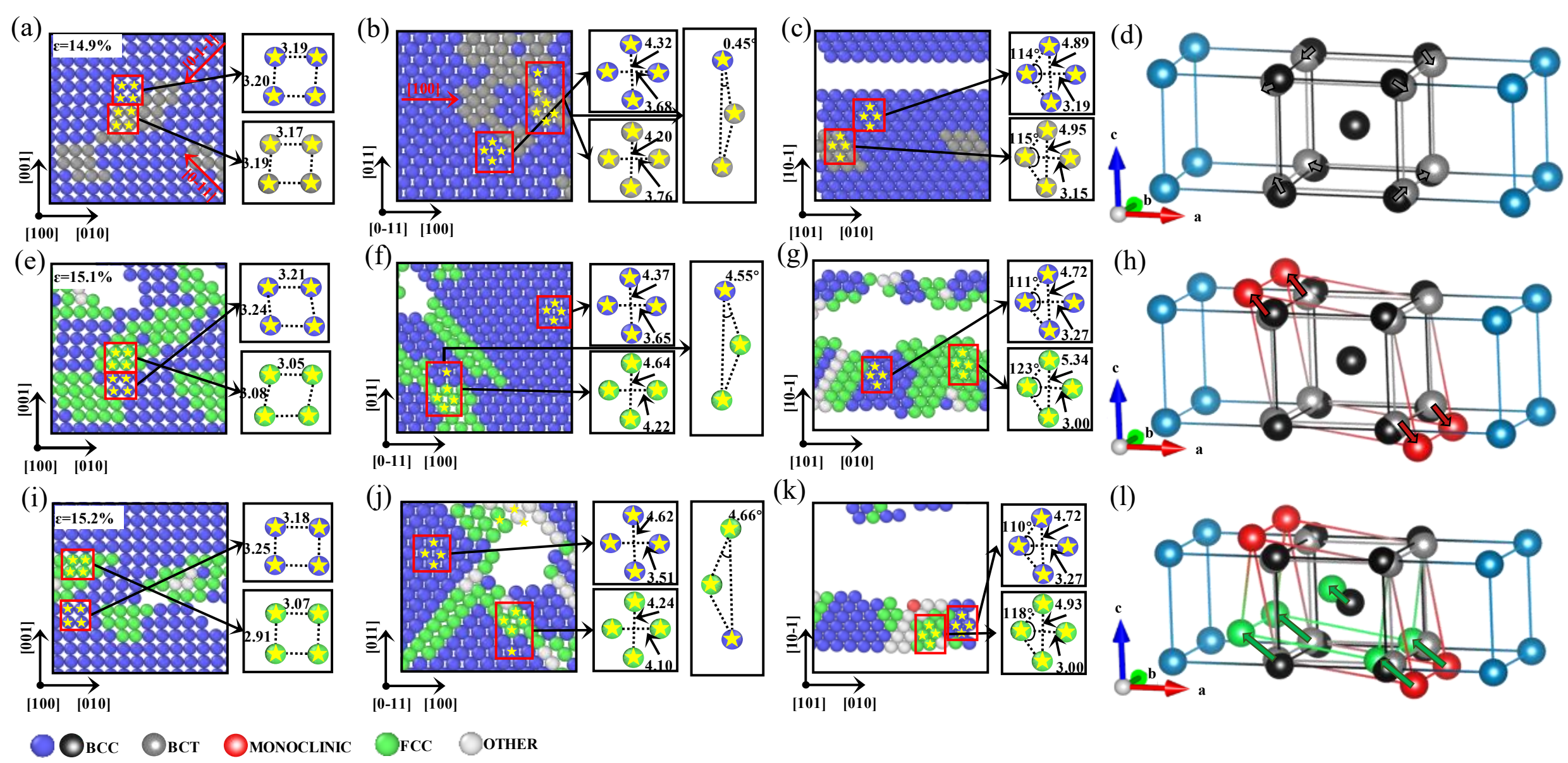

FIG 5. (a–c) At 1K bond length and bond angle relationships between BCC → BCT phases on (100), $(0\bar{1}1)$, and (101) planes, respectively (the CNA function of OVITO classifies BCT as FCC structure (FigS. 4); (d) Schematic of BCC→BCT transformation at ε = 14.9%.(e–g) Bond length and bond angle relationships between BCT → monoclinic phases on (100), $(0\bar{1}1)$, and (101) planes, respectively; (h) shear of BCT $(0\bar{1}1)$ plane along <$\bar{1}01$>/<$10\bar{1}$> directions at ε = 15.1% for BCT → monoclinic transition; (i–k) Bond length and bond angle

*qiuhong.hu@ucas.ac.cn
†hzshao@wzu.edu.cn
‡dai@ucas.ac.cn

109°28′, and the bond angle of the $(111)_{FCC}$ plane of the FCC is 120° (Fig. 5(k)). Our simulations reveal that BCC→FCC transformation takes place along the path of $(110)_{BCC} \rightarrow (001)_{FCC}$. The complete process of the transformation can be expressed as: BCC→ BCT → monoclinic → FCC.

**Influence of dislocation type on the BCC-FCC phase transformation**

Fig. 6 illustrates the temperature-dependent stress-strain responses of samples with edge-$\bar{b} \parallel z$, edge-$\bar{b} \parallel x$, screw-$\bar{b} \parallel x$, and screw-$\bar{b} \parallel z$ dislocations under uniaxial tension. Upon relaxation, dislocations remain stable at 1 K. However, at 300 K, 1000 K and 2000 K, all dislocations vanish, except for edge- $\bar{b} \parallel x$ (Fig. 6(b)). The analysis of the behavior of phase transformation on dislocation types revealed in the following. Stress accumulates near dislocations, regardless of the type of dislocation and temperature. Phase transformation preferentially initiates on {110} slip planes.

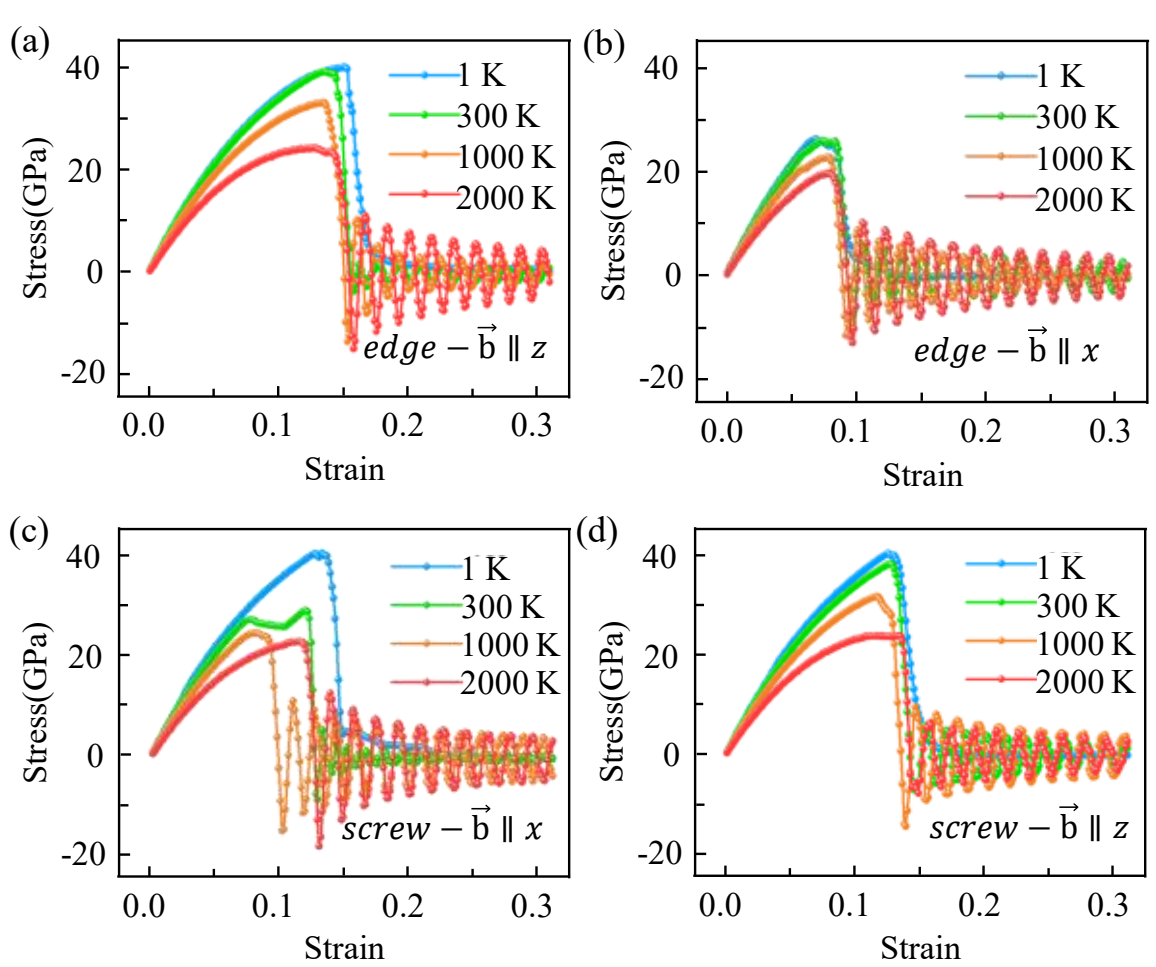

FIG 6. (a)(b)(c)(d) Stress-strain curves of single-crystal tungsten with edge-$\bar{b} \parallel z$ (a), edge-$\bar{b} \parallel x$ (b), screw-$\bar{b} \parallel x$ (c), and screw-$\bar{b} \parallel z$ (d) dislocations.

Edge-$\bar{b} \parallel z$ dislocations: at 1 K, the FCC phase fraction increases by almost double the amount as in the perfect crystal (FigS. 3(a)(e)), Dislocation cores along the tensile direction generate severe local stress, which readily activates {110} slip and triggers BCC → FCC phase transformation. However, the ratio of FCC to BCC at 300 K, 1000 K and 2000 K is similar to that in the perfect crystal at 1 K. This is mainly caused by the high mobility of dislocations. Notably, two ⟨100⟩ edge dislocations broke into thirty-three short ones during low-temperature tension, accompanied by a sharp increase in FCC phase content (FigS. 5(d)(e)). At higher temperatures, the type of slip planes is consistent with that in the perfect crystal in the intermediate to late stages.

Edge-$\bar{b} \parallel x$ dislocations: Less FCC forms than in the perfect crystal (Fig. 7(a)). Dislocation lines perpendicular to the loading axis greatly facilitate premature fracture. The low content of FCC phase indicates that the specimen fractures before accumulating sufficient strain to initiate phase transformation.

Screw- $\bar{b} \parallel x$ , and screw- $\bar{b} \parallel z$ dislocations: the dislocations remain stable during initial relaxation at 1 K. However, at 300 K, 1000 K and 2000 K, all dislocations vanish. For screw- $\bar{b} \parallel x$ case, as temperature increases, the FCC content decreases, from about 8% at 1 K to 1% at 2000 K. For screw-$\bar{b} \parallel z$ case, at all temperatures, the maximum content of FCC phase is still below 4% (Fig. 7(a)). This is probably because screw- $\bar{b} \parallel x$ dislocations induce more severe lattice distortion under tension along the x-direction and thus greatly facilitate the phase transformation.

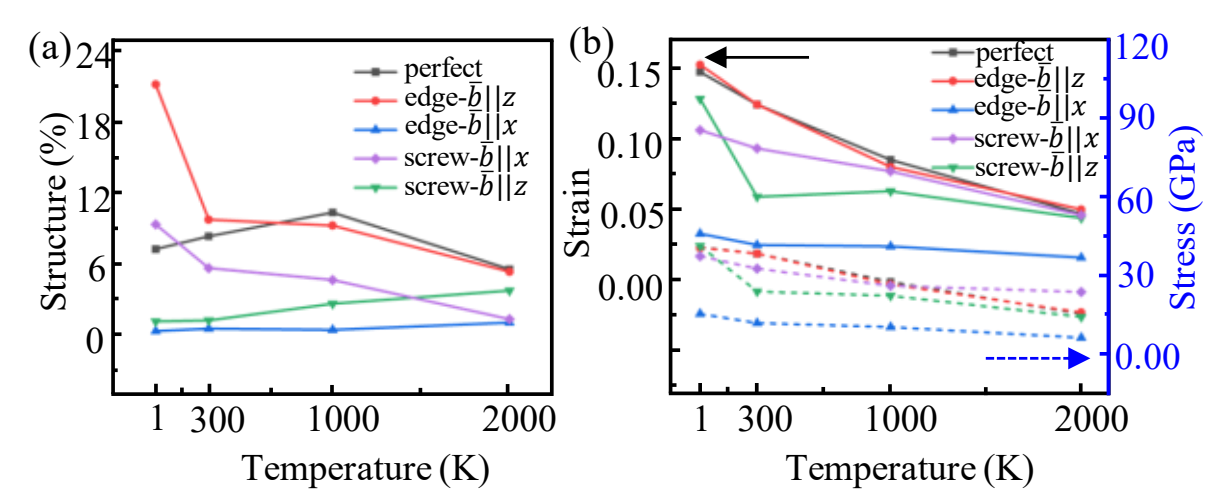

FIG 7. The results of the simulation on perfect-crystal system and systems containing edge-$\bar{b} \parallel z$, edge-$\bar{b} \parallel x$, screw-$\bar{b} \parallel x$, and screw-$\bar{b} \parallel z$ dislocations under tensile deformation at different temperatures: (a) peak FCC-phase fraction; (b) onset stress (dashed lines) and onset strain (solid lines) of phase transformation

Fig. 7 summarizes the overall effects of temperature, strain and dislocations on phase transformation. The type and orientation of dislocations can significantly modulate the tensile deformation behavior of materials. In the cases of edge-$\bar{b} || z$ and screw-$\bar{b} \parallel x$, the direction of the applied tensile stress is the same as the stress induced by the dislocation line and thus

*qiuhong.hu@ucas.ac.cn
†hzshao@wzu.edu.cn
‡dai@ucas.ac.cn

facilitates the BCC → FCC phase transformation. On the contrary, in the cases of edge-$\bar{b}||x$ and screw-$\bar{b} \parallel z$, the dislocation lines are perpendicular to the tensile axis. The stress field around the dislocation is perpendicular to the applied tensile stress, which triggers premature fracture. These differences in stress concentration and fracture behavior ultimately lead to pronounced distinctions in tensile mechanical curves and phase transformation evolution between the two sample systems.

**Conclusion**

In summary, we investigated the stress-strain relationship and corresponding microstructural evolution of single-crystal tungsten via GPUMD simulations, and found that uniaxial tensile loading can induce a BCC → FCC phase transformation. Temperature plays a decisive role in the phase transformation and plastic deformation. In the perfect crystal, the FCC content peaks at 10% at 1000 K (7.0% at 1 K, 8.3% at 300 K, and 6.0% at 2000 K), and the higher the temperature, the lower the strain required for the phase transformation. Pre-existing dislocations along the strain axis can promote the destabilization of the BCC structure, and facilitate the phase transformation. The maximum FCC content of 21.0% is achieved in the sample containing edge- $\bar{b}||z$ dislocation at 1 K. Slip predominantly occurs on {110} planes across all temperatures, and the crack tip exhibits Σ3 and Σ29 grain boundary characteristics. This phase transformation follows a three-step pathway of BCC→BCT→monoclinic phase → FCC. This study provides a mechanistic basis for strain-induced phase transformations in BCC metals and their alloys, and offers critical guidance for the experimental fabrication of high FCC content or duplex tungsten, addressing a longstanding challenge in tungsten material engineering for high-temperature structural applications.

We express our gratitude to Zhe-Yong Fan and Ke-Ke Song for their assistance in the training of machine learning potentials. We also extend our thanks to Yue-Hao Wang for his technical help in molecular dynamics simulations. Funding not applicable.

The authors declare no competing interests.

Data availability: The data that support the findings of this article are publicly available in Zenodo at Ref. [67].

Qiu-Hong Hu conceived the idea, supervised the study and participated in writing the paper. Yue-Yue Wang and Xin-Yi Leng conduct the DFT calculations and MD simulations, structural analysis, mechanical property evaluation, and data analysis. Yue-Yue Wang and Xin-Yi Leng also participated in writing the paper. He-Zhu Shao guided the MD simulations. Lu-Ru Dai directed the project. All authors participated in discussing the research results.

*qiuhong.hu@ucas.ac.cn
†hzshao@wzu.edu.cn
‡dai@ucas.ac.cn

*qiuhong.hu@ucas.ac.cn
†hzshao@wzu.edu.cn
‡dai@ucas.ac.cn

* qiuhong.hu@ucas.ac.cn
† hzshao@wzu.edu.cn
‡ dai@ucas.ac.cn